\documentclass{vldb}
\usepackage{graphicx}
\usepackage{balance}  
\usepackage{subcaption}
\usepackage{color}
\usepackage{booktabs}
\usepackage{comment}
\usepackage{xspace}
\usepackage{amsmath}
\usepackage[dvipsnames]{xcolor}
\usepackage{hyperref}
\usepackage[capitalise]{cleveref}

\PassOptionsToPackage{hyphens}{url}\usepackage{hyperref}
\DeclareRobustCommand{\subhead}[1]{\noindent\textbf{#1}} 

\renewcommand{\vec}[1]{\mathbf{#1}}
\DeclareMathOperator*{\argmax}{argmax}  

\def\sato{\mbox{\sc Sato}\xspace}
\def\base{\mbox{\sc Base}\xspace}
\def\wordfeature{\mbox{\sc Word}\xspace}
\def\charfeature{\mbox{\sc Char}\xspace}
\def\parafeature{\mbox{\sc Para}\xspace}
\def\statfeature{\mbox{\sc Stat}\xspace}
\def\satoLDA{${\text{\mbox{\sc Sato}}}_{\text{\tiny {\sc noStruct}}}$\xspace}
\def\satoCRF{${\text{\mbox{\sc Sato}}}_{\text{\tiny {\sc noTopic}}}$\xspace}
\def\Fone{$\text{F}_1$\xspace}
\def\sFone{support-weighted $\text{F}_1$\xspace}
\def\mFone{macro average $\text{F}_1$\xspace}
\def\SFone{Support-weighted $\text{F}_1$\xspace}
\def\MFone{Macro average $\text{F}_1$\xspace}

\def\tableVocab{table values\xspace}
\newcommand{\semantic}[1]{\texttt{#1}}
\newcommand{\atomic}[1]{\texttt{#1}}
\newcommand{\NAME}{\mbox{\sc Sato}}

\def\mispred{\textcolor{red}}
\def\crtpred{\textcolor{ForestGreen}}

\def\SHERLOCK{Sherlock\xspace}

\newcommand\mscript[1]{\mbox{\scriptsize\ensuremath{#1}}}

\newcommand{\new}[1]{{#1}}
\DeclareRobustCommand{\shepherding}[1]{#1}

\newcommand{\reminder}[1]{[#1]}

\newcommand{\yoshi}[1]{\reminder{{\bf\small\color{purple} (Yoshi)~#1}}}

\newcommand{\jinfeng}[1]{\reminder{{\bf\small\color{orange} (Jinfeng)~#1}}}

\makeatletter
\def\@copyrightspace{\relax}
\makeatother

\begin{document}


\title{Sato: Contextual Semantic Type Detection in Tables}



%
%
%
%

\numberofauthors{6} 

\author{
%
%
\alignauthor Dan Zhang\titlenote{Work done during internship at Megagon Labs}\\
       \affaddr{UMASS Amherst}\\
       \email{dzhang@cs.umass.edu}
\alignauthor Yoshihiko Suhara\\
       \affaddr{Megagon Labs}\\
       \email{yoshi@megagon.ai}
\alignauthor Jinfeng Li\\
       \affaddr{Megagon Labs}\\
       \email{jinfeng@megagon.ai}
\and  
\alignauthor Madelon Hulsebos\\
       \affaddr{The HEINEKEN Company}\\
       \email{mmhulsebos@gmail.com}
\alignauthor \c{C}a\u{g}atay Demiralp\\
       \affaddr{Megagon Labs}\\
       \email{cagatay@megagon.ai}
\alignauthor Wang-Chiew Tan\\
    \affaddr{Megagon Labs}\\
    \email{wangchiew@megagon.ai}
}

\maketitle
\begin{abstract}
Detecting the semantic types of data columns in relational tables is important for various data preparation and information retrieval tasks such as data cleaning, schema matching, data discovery, and semantic search. However, existing detection approaches either perform poorly with dirty data, support only a limited number of semantic types, fail to incorporate the table context of columns or rely on large sample sizes for training data. 
We introduce \sato, a hybrid machine learning model to automatically detect the semantic types of columns in tables, exploiting the signals from the context as well as the column values. \sato combines a deep learning model trained on a large-scale table corpus with topic modeling and structured prediction to achieve support-weighted and macro average F1 scores of 0.925 and 0.735, respectively, exceeding the state-of-the-art performance by a significant margin. We extensively analyze the overall and per-type performance of \sato, discussing how individual modeling components, as well as feature categories, contribute to its performance. 
\end{abstract}
\section{Introduction}\label{sec:intro}

Many data preparation and information retrieval tasks including data cleaning, integration, discovery and search rely on the ability to accurately detect data column types. Automated data cleaning uses transformation and validation rules that depend on data types~\cite{2011-wrangler,Raman:2001:PWI:645927.672045}. 
Schema matching for data integration leverages data types to find correspondences between data columns across tables~\cite{rahm2001survey}. Similarly, data discovery benefits from detecting the types of data columns in order to return semantically relevant results for user queries~\cite{aurum,seeping-semantics}. Recognizing the semantics of table values helps aggregate information from multiple tabular data sources. Search engines also rely on the detection of semantically relevant column names to extend support to tables~\cite{venetis2011recovering}.  

We can consider two categories of types for table columns:  atomic and semantic. Atomic types such as \atomic{boolean}, \atomic{integer}, and \atomic{string} provide basic, low-level type information about a column. On the other hand, semantic types  such as \semantic{location}, \semantic{birthDate}, and \semantic{name}, convey finer-grained, richer information about column values. Detecting semantic types can be a powerful tool, and in many cases may be essential for enhancing the effectiveness of data preparation and analysis systems. In fact, commercial systems such as Google Data Studio \cite{googledatastudio}, Microsoft Power BI~\cite{powerbi}, Tableau~\cite{tableau}, and Trifacta~\cite{trifacta} attempt to detect semantic types, typically  using  a combination of regular expression matching and dictionary lookup.  While reliable for detecting atomic types and simple, well-structured semantic types such as credit card numbers or e-mail addresses,  these rule-based approaches are not robust enough to process dirty or missing data, support only a limited variety of types, and fall short for types without strict validations. However, many tables found in legacy enterprise databases and on the Web have column names that are either unhelpful (cryptic, abbreviated, malformed, etc.) or missing altogether.  

\begin{figure*}[tbh]
    \centering
    \includegraphics[width=\linewidth]{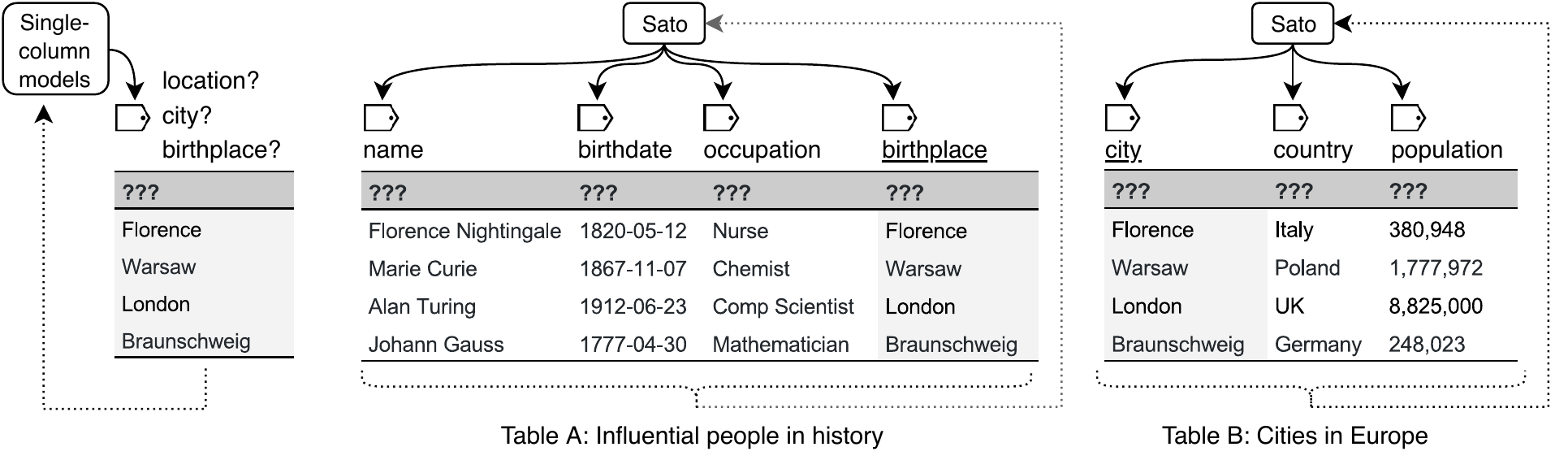}
    \caption{Two actual tables with unknown column types (Table A and Table B) from the VizNet corpora.  The last column of Table A and the first column of Table B  have identical values: `Florence,' `Warsaw,' `London,' and `Braunschweig.' However powerful, a prediction model based solely on column values (i.e., single-column prediction) cannot resolve the ambiguity to infer the correct semantic types, \semantic{birthplace} and \semantic{city}. \sato incorporates signals from table context and perform a multi-column type prediction to help effectively resolve ambiguities like these and improve the accuracy of semantic type predictions. \label{fig:start-example}
    \vspace{-10pt}
   }
    
\end{figure*}
In response, recent work~\cite{Hulsebos:2019:KDD} introduced Sherlock, a deep learning model for semantic type detection trained on a massive table corpora ~\cite{hu2019viznet}.  Sherlock formulates semantic type detection as a multi-class classification problem where classes correspond to semantic types. It leverages more than 600K real-world table columns for learning with a multi-input feed forward deep neural network, providing state-of-the-art results. 

While Sherlock represents a significant leap in applying deep learning to semantic typing, it suffers from two problems. First,  it under-performs for types that do not have a sufficiently large number of samples in the training data.  Although this is a known issue for deep learning models, it nevertheless restricts Sherlock's application to underrepresented types, which form a long tail of data types appearing in tables at large. 
Second, Sherlock uses only the values of a column to predict its type, without considering the column's context in the table. Predicting the semantic type of a column based solely on the column values, however, is an under-determined problem in many cases. 

Consider the example in \cref{fig:start-example}: for a column that contains `Florence,' `Warsaw,' `London,' and `Braunschweig' as values, \semantic{location}, \semantic{city}, or \semantic{birthPlace} could all be reasonable semantic types for the column. It can be hard to resolve such ambiguities using only column values because the semantic types also depend on the \textit{context} of the table. Continuing with the example, it is highly likely that the column's type would be \semantic{birthPlace} if it came from Table A since the table contains biographical information about influential personalities. However, the same column in Table B would be more likely to have the type \semantic{city}, as the table's other columns present information about European cities. 

In this paper, we introduce {\bf \sato} ({\bf S}em{\bf A}ntic {\bf T}ype detection with table c{\bf O}ntext), a hybrid machine learning model that incorporates table contexts to predict the semantic types of table  columns. 
\sato combines topic modeling~\cite{Blei:2012:TopicModels} and structured learning~\cite{Lafferty:2001:CRF} together with single-column type prediction based on the Sherlock model. Similar to earlier work~\cite{Hulsebos:2019:KDD}, we consider 78 common semantic types and use the WebTables dataset from the VizNet corpus~\cite{hu2019viznet} to train our model. 

We summarize our main contributions below:
\shepherding{\begin{enumerate}
    \item \sato significantly outperforms the state-of-the-art in semantic type prediction, increasing the macro and support-weighted F1 scores by as much as 14.4\% and 5.3\%, respectively. Through a comparative analysis of per-type predictions, we also show 
    that \sato's performance gains are primarily due to improved predictions for underrepresented semantic types in the long tail. To facilitate future research and applications, we have released \sato as an open source project along with an online demo at \url{https://github.com/megagonlabs/sato}.
    \item \sato achieves this high prediction accuracy using a novel hybrid model that regulates semantic type prediction using signals from the global context (values from the entire table) and the local context (predicted types of neighboring columns,) demonstrating the effectiveness of incorporating table context into semantic type detection. 
    \item \sato also introduces a new extensible architecture for type detection with modules for modeling single columns, global context, and local context. One can easily plug in a different single-column model while keeping the rest intact. We demonstrate this extensibility in~\cref{sec:learned-rep} by replacing the default single-column predictor 
    in \sato with BERT~\cite{devlin2019bert}.
\end{enumerate}}

\section{Problem Formulation}\label{sec:prelim}
Our goal is to predict semantic types for table columns using their values, without considering the header information. 
We formulate it as multi-class classification, each class corresponding to a predefined semantic type.


We consider the training data as a set of tables. 
Let $c_1, c_2,\ldots, c_m$ be the columns of 
a given table and $t_1, t_2, \ldots, t_m$ be the true semantic types of these columns, where $t_i\in\mathcal{T}$, the set of labels for possible semantic types considered (e.g., \semantic{city}, \semantic{country}, \semantic{population}).  
Similarly, let $\Phi$ be a feature extractor function that takes a single column $c_i$ and returns an $n$-dimensional feature vector $\Phi_{i}$. 
One approach to semantic typing is to learn a mapping $f_{\rm single}:{\Phi}^n \rightarrow \mathcal{T}$ from values of single columns to semantic types. We refer to this model as {\it single-column prediction}. The Sherlock~\cite{Hulsebos:2019:KDD} model falls into this category.

In \sato, in order to make the best use of table contexts and resolve semantic ambiguity with single-column predictions, we formulate the problem as {\it multi-column prediction}. A multi-column prediction model learns a mapping $f_{\rm mult}: {\Phi}^{n\times m} \rightarrow \mathcal{T}^m$ from the entire table (a sequence of columns) to a sequence of semantic types. This formulation allows us to incorporate table context into semantic type prediction in two ways. 

First, we use features generated from the entire table as table context. For example, the column values `Italy,' `Poland,' ... and  `380,948,' `1,777,972,' ... are also used to predict the semantic type of the first column in Table B (in \cref{fig:start-example}.) Second, we can jointly predict the semantic types of columns from the same table. Again, for Table B, with the joint prediction the predicted types \semantic{country} and \semantic{population} of neighboring columns would help to make a more accurate prediction for the first column.

\section{Model}\label{sec:model}
\noindent
\shepherding{\subhead{Table context} As demonstrated in~ \cref{fig:start-example}, the contextual information of a column can be used to resolve ambiguities and improve the semantic type prediction for the column. To this end, we identify two basic types of context that collectively characterize the context of a table column: global context and local context.  We define the global context for a column to be the set of all the cell values in the table.  In this sense, all the columns in a given table have the same global context.  We show in~\cref{sec:model-LDA} how the global context can be used to compute a global descriptor effectively capturing the intent of a table.  
We define the local context of a column as the set of independently predicted semantic types of the neighboring columns in the same table.  Local context can be used to resolve semantic type ambiguities when combined with single-column predictions. The scope of such a local neighborhood is flexible and can be adjusted based on the desired trade-off between model performance and model complexity. In this work, we restrict the local neighborhood to immediately adjacent columns. We demonstrate in \cref{sec:model-structured} how local context can be effectively used to improve the semantic type detection accuracy through structured predictions. 
}

\begin{figure}[ht]
    \centering
    \includegraphics[width=0.45\textwidth]{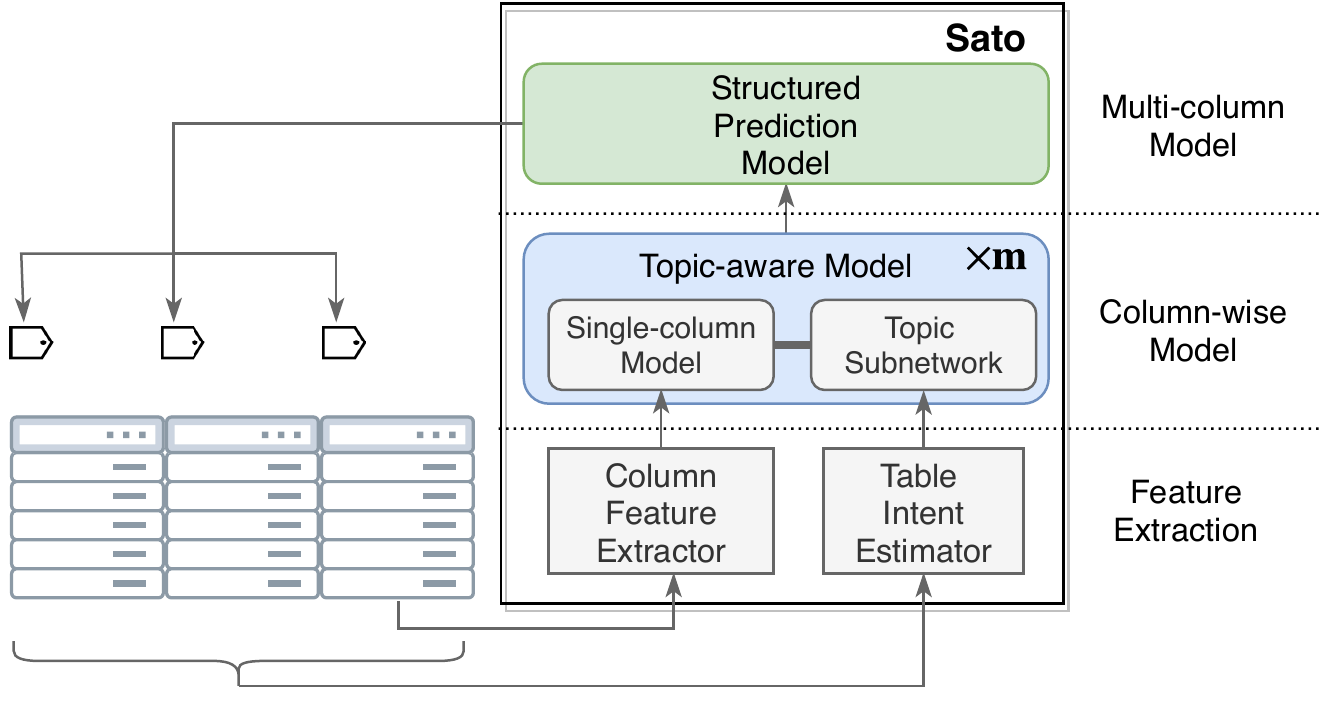}
    \caption{In \sato, the \textit{topic-aware} module extends single-column models with additional topic subnetworks, incorporating a context modeling table intent into the model. The \textit{structure prediction} module then combines the topic-aware results for all $m$ columns, providing the final semantic type prediction for the columns in the table. \label{fig:overview}
    \vspace{-10pt}}
\end{figure}

\shepherding{
Next, we will show how \sato effectively captures contextual signals from both global and local sources using a hybrid machine learning model.} 
It has two modeling components: (1) A topic-aware prediction component that estimates the {\em intent} (a global descriptor) of a table using topic modeling and extends the single-column prediction model with an additional topic subnetwork.
(2) A structured output prediction model that  combines the topic-aware predictions for all $m$ columns and performs multi-column joint semantic type prediction.  \cref{fig:overview} illustrates the high-level architecture of \sato.  We next discuss each \sato component and its implementation in detail. 


%

\subsection{Single-column prediction model}\label{sec:model-single-column}

As shown in \cref{fig:overview}, \sato's topic-aware module is built on top of a single-column prediction model that uses a deep neural network. We first provide a brief background on deep learning and a description of the single-column model.   
\\
\\
\subhead{Deep learning} Deep learning~\cite{LeCun:2015:DeepLearning} is a form of representation learning that uses neural networks with multiple layers. Through simple but non-linear transformations of the input representation at each layer, deep learning models can learn representations of the data at varying levels of abstractions that are useful for the problem at hand (e.g., classification, regression). 
%
Deep learning combined with the availability of massive table corpora~\cite{webtables,hu2019viznet} presents opportunities to learn from tables in the wild~\cite{halevy2009unreasonable}. It also presents opportunities to improve existing approaches to semantic type detection as well as other research problems related to data preparation and information retrieval.  Although prior research has used shallow neural networks for related tasks (e.g.,~\cite{li1994semantic}), it is only more recently that Hulsebos et al.~\cite{Hulsebos:2019:KDD} developed Sherlock, a large-scale deep learning model for semantic typing.
\\
\\
\subhead{Deep learning for single-column prediction}
\sato builds on single-column prediction by using column-wise features and employs an architecture which allows any single-column prediction model to be used. In our current work, we choose Sherlock as our single-column prediction model due to its recently demonstrated performance. 

The column-wise features used in \sato include character embeddings (\charfeature), word embeddings (\wordfeature), paragraph embeddings (\parafeature), as well as column statistics (e.g., mean, std) (\statfeature.) 

A multi-layer subnetwork is applied to the column-wise features to compress high-dimensional vectors into compact dense vectors, with the exception of the \statfeature feature set, which consists of only 27 features. The output of the three subnetworks is concatenated to the statistical features, forming the input to the primary network. After the concatenation of these features, in the primary network two fully-connected layers (ReLU activation) with BatchNorm and Dropout layers are applied before the output layer. The final output layer, which includes a softmax function, generates confidence values (i.e., probabilities) for the 78 semantic types.

\begin{figure*}[ht]
\begin{subfigure}[b]{0.5\linewidth}
\centering
\includegraphics[width=0.5\textwidth]{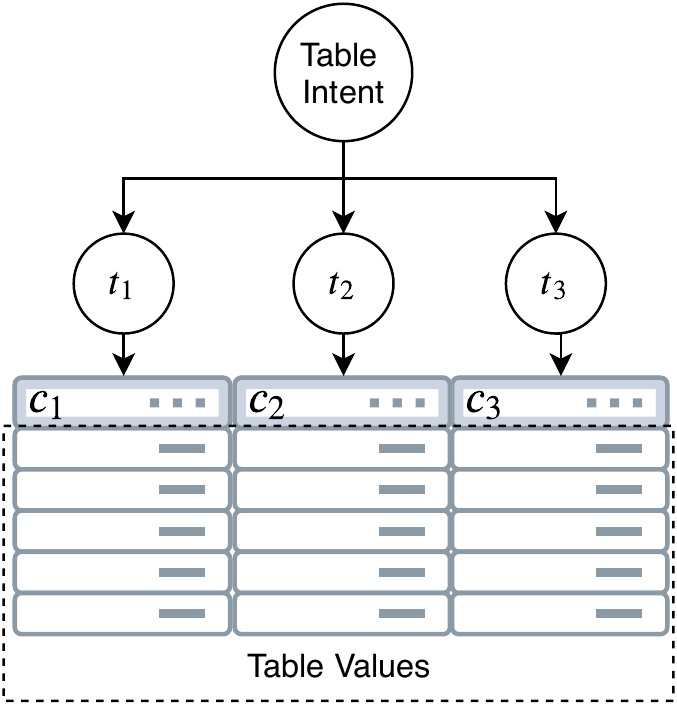}
\caption{\label{fig:LDA_GM}}
\end{subfigure}%
\begin{subfigure}[b]{0.5\linewidth}
\centering
\includegraphics[width=0.7\textwidth]{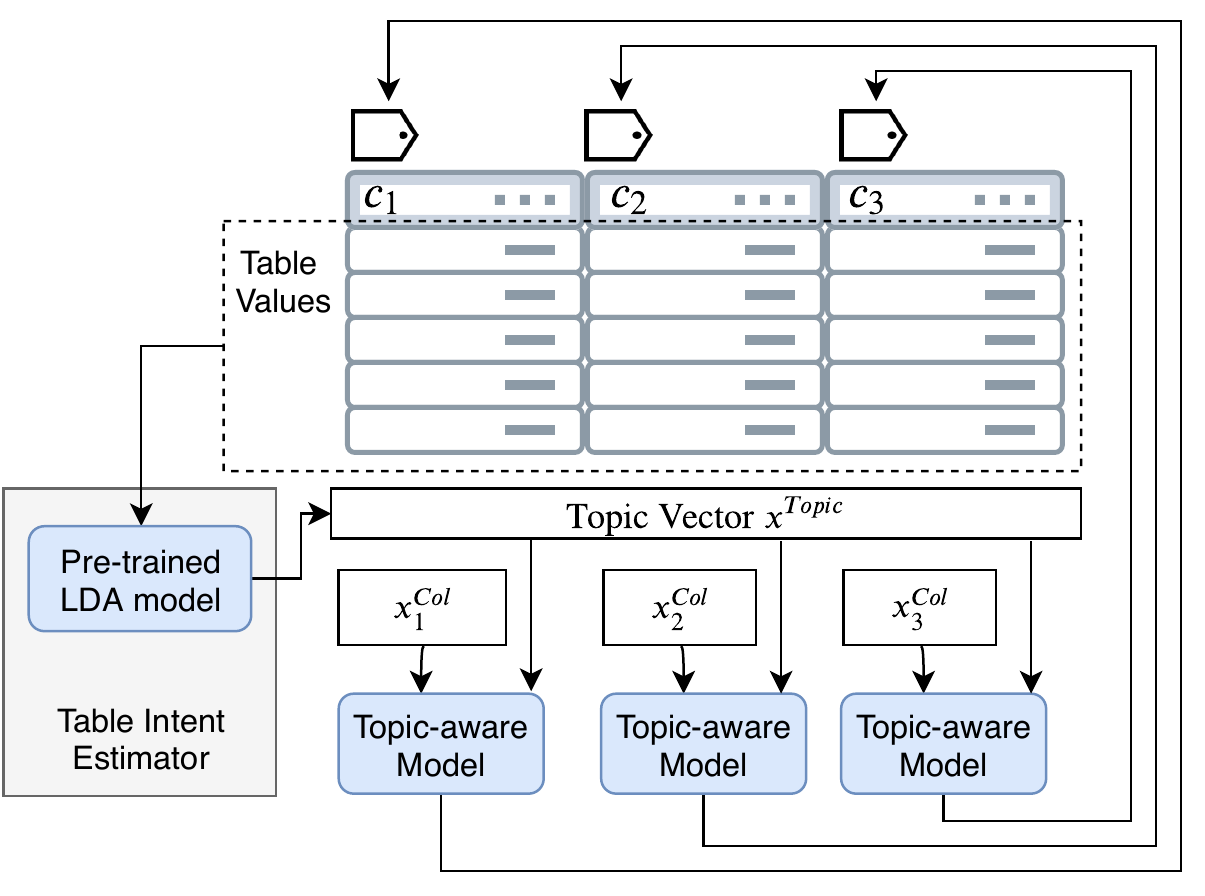}
\caption{\label{fig:LDA_prediction}}
\end{subfigure}
\caption{\sato's topic-aware modeling is based on the premise that every table is created with an {\em intent} in mind and that the semantic types of the columns in a table are expressions of that intent with thematic coherence.  In other words, (a) the intent of a table determines the semantic types of the columns in the table, which in turn generate the column values,  acting as latent variables. (b) \sato estimates the intent of a given table with a topic vector obtained from a pre-trained LDA model and combines it with the local evidence from per-column values using a deep neural network.}
\end{figure*}

\subsection{Topic-aware prediction model}\label{sec:model-LDA}
The first component of \sato is a topic-aware prediction module. \shepherding{This module first creates a vector representation for the global context of a given table by computing a topic vector from the values of the entire table.} The topic-aware prediction module then feeds this topic vector as input to the column-wise prediction model. The column-wise prediction model extends the neural network model above with an additional subnetwork in order to take topic 
vectors as input. 
We next discuss how taking the global context of a table into account 
in  semantic type prediction can help resolve ambiguities. 

%
%

\subhead{Table semantics}  Tables are collections of related data entities organized in rows. \new{To incorporate table semantics in our model,  we build on an intuition by Venetis et al.~\cite{venetis2011recovering} that  a user constructing a table has a particular \textit{intent} or schema in mind.  We extend this intuition and argue that semantic types of the columns in a table can be considered a meaningful expression (or utterance) of the user intent.}  Each column of the table partially fulfills the intent by describing one attribute of the entities. As illustrated in \cref{fig:LDA_GM}, \new{the intent of a table is a latent component determining the semantic types of the columns in the table, which in turn generates the column values.} We refer to the set of all column values in a table as \textit{table values}.

Thus, being able to accurately infer the table intent can help to improve the prediction of column semantics. Table captions or titles usually capture table intent. For example, in \cref{fig:start-example}, Table A intends to provide biographical information about influential personalities in history and Table B talks about geographical information about cities in Europe. However, as with column semantics, a clear and well-structured description of intent is not always available in real-world tables. Therefore we need to estimate table intent without relying on any header or meta information.

\sato estimates a table's intent by mapping its values onto a low-dimensional space. Each of these dimensions 
corresponds to a ``topic,'' describing  one aspect of a possible table intent. The final estimation is a distribution over the latent topic dimensions generated using topic  modeling approaches. Next, we provide a brief background on topic models and explain how \sato extracts topic vectors from tables and 
feed them to topic-aware models.
\\
\\
\subhead{Topic models} Finding the topical composition of textual data is useful for many tasks, such as document summarization or featurization. Topic models~\cite{Blei:2012:TopicModels} aim to automatically discover thematic topics in text corpora and discrete data collections in an unsupervised manner. Latent Dirichlet allocation (LDA)~\cite{Blei:2003:LDA} is a simple yet powerful generative probabilistic topic model, widely used for quantifying thematic structures in text.  LDA represents documents as random mixtures of latent topics and each latent topic as a distribution over words. The main advantage of probabilistic topic models such as LDA over clustering algorithms is that probabilistic topic models can represent a data point (e.g., document) as a mixture of topics. 
Although LDA was originally applied to text corpora, since then many variants have been developed to discover thematic structures in non-textual data as well (e.g.,~\cite{blei2003modeling,fei2005bayesian,Yuan:2012:DRD}.)
\\
\\
\subhead{Table intent estimator}  We use an LDA model to estimate a table's intent as a topic-vector, treating values of each table as a ``document.''  As illustrated in \cref{fig:LDA_prediction}, we implement the table intent estimator as a pre-trained LDA model. It takes table values as input and outputs a fixed-length vector named ``table topic vector'' over the topic dimensions. For \sato, we pre-train an LDA model with 400 topic dimensions on public tables that have had their headers and captions removed. 

The topics are generated during training according to the data's semantic structure, so they do not have pre-defined meanings. However, by looking at the representative semantic types associated with each topic, we found some examples with good interpretations. For example, topic \# 192 is closely associated with the semantic types ``origin, nationality, country, continent, and sex'' and thus possibly captures aspects about personal information, while topic \# 264 corresponds to ``code, description, create, company, symbol'' and can be interpreted as a business-related topic. Detailed topic analysis can be found in \cref{sec:topic_analysis}.
\\
\\
\subhead{Learning and prediction} \cref{fig:LDA_prediction} shows how topic-aware models take the values in a table topic vector as additional features for both learning and prediction. We augment the single-column neural network model with an additional subnetwork to take topic vectors as input and then append its output before feeding into the primary network. In this way, the topic-aware model will learn not only relationships between the input column and its type but also how the column type correlates to the table-level contextual information.

\begin{figure*}[ht]
\begin{subfigure}[b]{0.5\linewidth}
\centering
\includegraphics[width=0.5\textwidth]{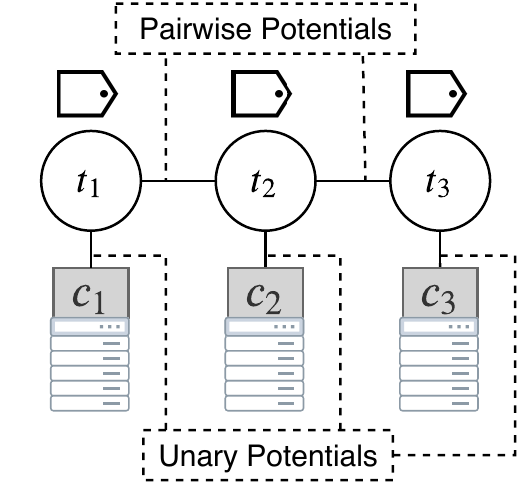}
\caption{\label{fig:CRF_model}}
\end{subfigure}%
\begin{subfigure}[b]{0.5\linewidth}
\centering
\includegraphics[width=0.8\textwidth]{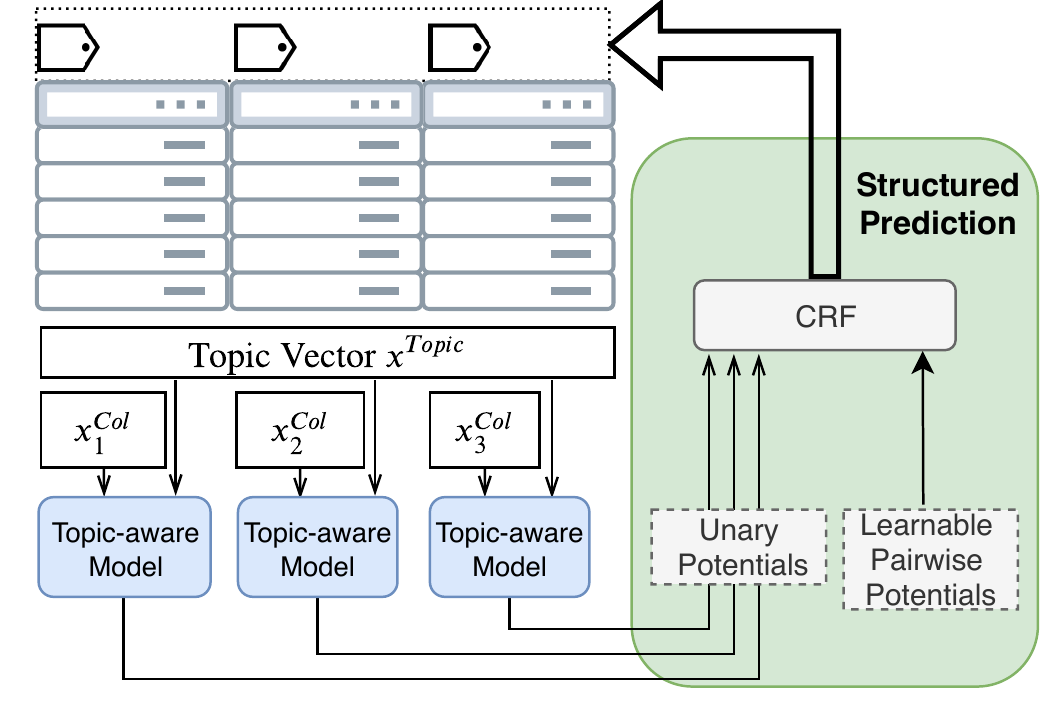}
\caption{\label{fig:CRF_pred}}
\end{subfigure}
\caption{(a) \sato uses a linear-chain CRF to model the dependencies between columns types given their values.
    (b) For each column, \sato plugs in the column-wise prediction scores for each type as the unary potentials of the corresponding node in the CRF model. Then \sato learns the pairwise potential through backpropagation updates using stochastic gradient descent, maximizing the posterior probability  $P(\vec{t}|\vec{c})$.  
     Although we choose to use predictions from topic-aware models in the current implementation, the \sato architecture is flexible to support unary potentials from arbitrary column-wise models.}
 \vspace{-3mm}
\end{figure*}

\subsection{Structured prediction model\label{sec:model-structured}} 


\shepherding{We have described how \sato captures the global context of a column by computing a topic vector for the entire table and passing it to the single-column model as additional input. Incorporating only the global context into the model may not be sufficient, however, as the topic-aware model does not directly take the relationships among the semantic types of neighboring columns (i.e., local context) into account. Therefore, we incorporate local context using a structured output prediction model which comprises the second component of \sato.}

Through preliminary analysis, we confirm that certain pairs of semantic types co-occur in tables much more frequently than others. For example, in a WebTables sample, the most frequent pair \semantic{city} and \semantic{state} co-occurs 4 times more often than the tenth most frequent pair \semantic{name} and \semantic{type} (detailed co-occurrence statistics available in \cref{sec:dataset}). Such inter-column relationships show the value of ``local'' contextual information from surrounding columns in addition to the ``global'' table topic. \sato models the relationships between columns through pairwise dependencies in a graphical model and performs table-wise prediction using structured learning techniques.
\shepherding{Although the notion of local context is not limited to immediate neighbors, \sato only models pairwise relations between adjacent columns because of its simplicity, efficiency, and empirical accuracy. We leave the study of the broader local context, which can be modeled using high-order graphical models (further discussed in \cref{sec:high_order_CRF}), as future work.}

%
%

\subhead{Structured output learning} In addition to semantic type detection, many other prediction problems such as named entity extraction, language parsing, and image segmentation have spatial or semantic structures that are inherent to them. Such structures mean that predictions of neighboring instances correlate to one another. Structured learning algorithms~\cite{bakir2007predicting}, including probabilistic graphical models~\cite{koller2009probabilistic} and recurrent neural networks~\cite{lstm97hochreiter, rumelhart1986learning}, model dependencies among the values of structurally linked variables such as neighboring pixels or words to perform joint predictions.  

A conditional random field (CRF)~\cite{Lafferty:2001:CRF} is a discriminative undirected probabilistic graphical model and a popular techniques for structured learning with successful applications in labeling, parsing and segmentation problems across domains. Similar to Markov random fields (MRFs)~\cite{geman1986markov, koller2009probabilistic}, exact inference for general CRFs is intractable but there are special structure such as linear-chains that allow exact inference.  There are also several efficient approximate inference algorithms based on message passing, linear-programming relaxation, and graph cut optimization for CRFs with general graphs~\cite{Lafferty:2001:CRF}. 
\\
\\
\subhead{Modeling column dependencies}  \sato uses a linear-chain CRF to explicitly encode the inter-column relationship while still considering features for each column.  
%
We encode the output of a column-wise prediction model (i.e., predicted semantic types of the columns) and the combinations of semantic types of columns in the same table as CRF parameters.
As shown in \cref{fig:CRF_model}, in the CRF model, each variable $t_i$ represents the type of a column with corresponding column values $c_i$ as the observed evidence. Variables representing the types of adjacent columns are linked with an edge. Given a {\it sequence} of columns $\vec{c}$ in a table, the goal is to find the best {\it sequence} of semantic types $\vec{t}$, which provides the largest conditional probability $P(\vec{t}|\vec{c})$.
%

%
The conditional probability can be written as a normalized product of a set of real-valued functions. Following the convention, we refer to these functions in log scale as ``potential functions.'' \textit{Unary potential} $\psi_{\rm UNI}(t_i, c_i)$ captures the likelihood of predicting type $t_i$ based on the content of the corresponding column $c_i$. 
\textit{Pairwise potential} $\psi_{\rm PAIR}(t_i, t_j)$ represents the  ``coupling degree'' between types $t_i$ and $t_j$. 
%

We use a linear-chain CRF, where the conditional distribution is defined by the unary prediction potentials and pairwise potentials between adjacent columns:
\begin{gather}
\scalebox{0.93}{$
 \begin{aligned}
 P(\vec{t}|\vec{c}) = \frac{1}{Z(\vec{c})} \exp\left(\sum_{i=1}^{m} \psi_{\rm UNI}(t_i, c_i) + \sum_{i=1}^{m} \sum_{j=i+1}^{m} \psi_{\rm PAIR}(t_i, t_j)\right),\notag
 \end{aligned}
$}
\end{gather}
where 
\begin{gather}
\scalebox{0.93}{$
 \begin{aligned}
Z(\vec{c}) = \sum_{\vec{t}} \exp\left(\sum_{i=1}^{m}\psi_{\rm UNI}(t_i, c_i) + \sum_{i=1}^{m}\sum_{j=i+1}^{m} \psi_{\rm PAIR}(t_i, t_j)\right) \notag
\end{aligned} $}
\end{gather}
is an input-dependent normalization function.
\\
\\
\subhead{Unary potential functions}
We use unary potentials to model the probability of a semantic type given the column content. In other words, the unary potential of 
a semantic type for a given column can be considered the probability of that semantic type based on the values of the column. 
The architecture of \sato supports using estimates of any valid  column-wise prediction model as unary potentials. In the current work,  we obtain the unary potentials of the semantic types for a given column from the output of our topic-aware prediction model, which uses both table-level topic vector and column features as input.
\new{Using the examples of \cref{fig:start-example}, we expect that the highlighted column, which contains `Florence,’ `Warsaw,' `London,’ and `Braunschweig’, would have high unary potential scores for location-related semantic types such as \semantic{location}, \semantic{city}, and \semantic{birthplace}. In other words, the unary potentials calculate column-wise prediction scores, which are used to {\it select} semantic type candidates for each column.}
\\
\\
\subhead{Pairwise potential functions}
Pairwise potentials capture the relationship between the semantic types of two columns in the same table. These relationships  can be parameterized with a $|\mathcal{T}|\times|\mathcal{T}|$ matrix $P$, where $\mathcal{T}$ is the set of all possible types and $P_{ij}$ ($= \psi_{\rm PAIR}(t_i, t_j)$) is a weight parameter for the ``coupling degree'' of semantic types $t_i$ and $t_j$ in adjacent columns. Such a coupling degree can be approximated by the co-occurrence frequency. We expect the pairwise weight of two semantic types to be proportional to their frequency of co-occurrence in adjacent columns.  Pairwise potential weights in our CRF model are trainable parameters, updated by gradient descent. 
%
\new{Through the training step, we expect that \sato updates the CRF parameters so that frequently co-occurred pairs like (\semantic{city}, \semantic{country}) and (\semantic{occupation}, \semantic{birthplace}) have higher pairwise potential scores. Thus, the trained model can resolve the disambiguate issue (shown in \cref{fig:start-example}) by using pairwise potentials and achieves context-aware predictions.}
\\
\\
\subhead{Learning and prediction}
We use the following objective function to train a \NAME{} model. The objective function is the log-likelihood of semantic types of columns in the same table:
\begin{gather}
\scalebox{0.93}{$
 \begin{aligned}
\log P(\vec{t}|\vec{c})= \sum_{i=1}^{m}\psi_{\rm UNI}(t_i, c_i) + \sum_{i=1}^{m} \sum_{j=i+1}^{m}\psi_{\rm PAIR}(t_i, t_j)-\log{Z(\vec{c})}. \notag
\end{aligned} $}
\end{gather}
Here, the normalization term $Z$ sums over all possible semantic type combinations. To efficiently calculate $Z$, we can use the forward-backward algorithm~\cite{rabiner1989tutorial}, which uses dynamic programming to cache intermediate values while moving from the first to the last columns.
After the training phase, as shown in \cref{fig:CRF_pred}, \sato performs holistic type prediction with learned pairwise potential and unary potential provided by topic-aware prediction. To obtain prediction results, we conduct maximum a posteriori (MAP) inference of semantic types:
\begin{gather}
\scalebox{0.93}{$
 \begin{aligned}
 \hat{\vec{t}} = \argmax_{\vec{t}} {\log P(\vec{t}|\vec{c})} = \argmax_{\vec{t}} \left(\sum \psi_{\rm UNI} + \sum\sum\psi_{\rm PAIR}\right). \notag
\end{aligned} $}
\end{gather}
$Z(\vec{c})$ does not affect $\argmax$ since it is a constant with respect to $\vec{t}$. Then we use the Viterbi algorithm~\cite{viterbi1967error} to calculate and store partial combinations with the maximum score at each step of the column sequence traversal, avoiding redundant computation.  

\section{Evaluation\label{sec:eval}}
We compare \sato and its two basic variants obtained by ablation with the state-of-the-art Sherlock~\cite{Hulsebos:2019:KDD} implemented as the \base method. We omit comparisons with matching-based algorithms, decision-tree-based semantic typing since they are outperformed by Sherlock as demonstrated in \cite{Hulsebos:2019:KDD}.
\begin{figure*}
\centering
\includegraphics[width=0.9\textwidth]{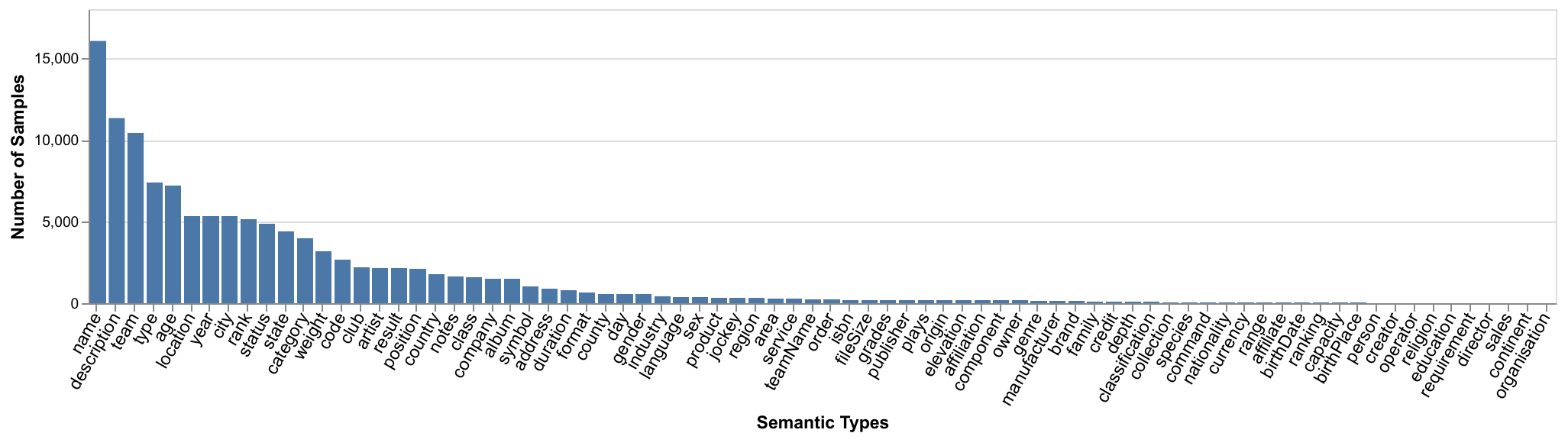}
\caption{Counts of the 78 semantic types in the dataset $\mathcal{D}$ form a long-tailed distribution. \sato improves the prediction accuracy for the types with fewer samples (those in the long-tail) by effectively incorporating table context. \label{fig:type-dist}}
\end{figure*}

\begin{figure}[h]
 \centering
\includegraphics[width=0.9\columnwidth]{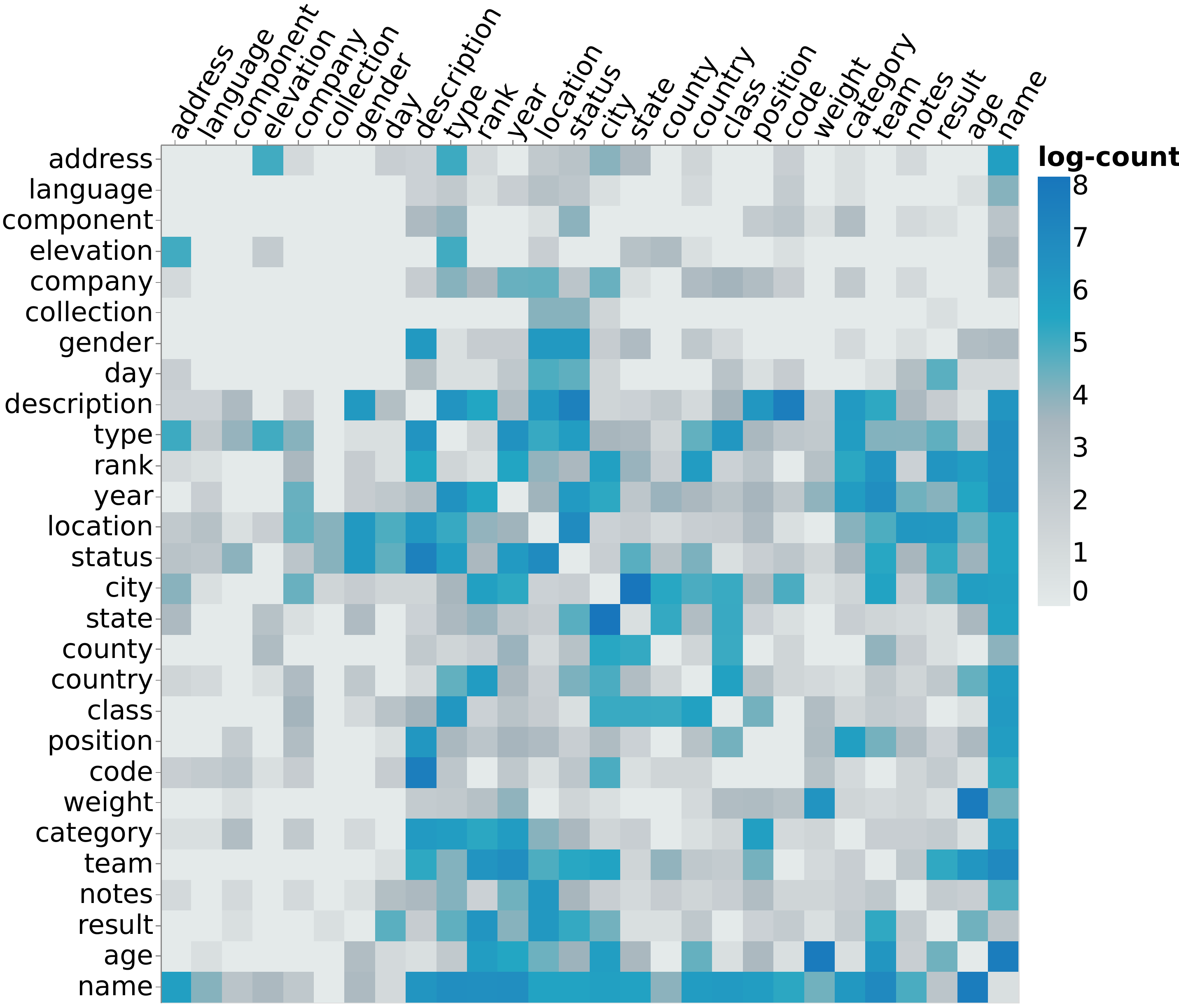}
\caption{Co-occurrence frequencies in log scale for a selected set of types. Certain pairs like (\semantic{city}, \semantic{state}) or (\semantic{age}, \semantic{weight}) 
appear in the same table more frequently than others. There are non-zero diagonal values as tables can have multiple columns of the same semantic type.\label{fig:co-occur}
}
\end{figure}

\subsection{Datasets\label{sec:dataset}}
We evaluate the effectiveness of the proposed models on the WebTables corpus from VizNet~\cite{hu2019viznet} and restrict ourselves to the relational web tables with valid headers that appear in the 78 semantic types. \new{These types resulted from a selection process~\cite{Hulsebos:2019:KDD} and originate from the T2Dv2 Gold Standard\footnote{\url{http://webdatacommons.org/webtables/goldstandardV2.html}}, which describes 237 DBpedia properties frequently occurring in the WebTables corpus. }
To avoid filtering out columns with slight variation in capitalization and representation, we convert all column headers to a ``canonical form'' before matching. The canonicalization process starts with trimming content in parentheses. We then convert strings to lower case, capitalize words except for the first (if there are more than one word) and concatenate the results into a single string. For example, strings `YEAR,' `Year' and `year (first occurrence)' will all have canonical form `year,' and `birth place (country)' will be converted to `birthPlace.'


Since we formulate semantic typing as a multi-column type detection problem, we extract 80K tables, instead of columns, from a subset of the WebTables corpus as our dataset $\mathcal{D}$. The column headers in their canonical forms act as the groundtruth labels for semantic types. To help evaluate the importance of incorporating table semantics, we also create a filtered version $\mathcal{D}_{mult}$ with 33K tables. We filter out singleton tables (those containing only one column) since they lack context as defined in this paper. 
\new{We then conduct 5-fold cross-validation where we use 80\% of the tables for training and a held-out set (20\%) for evaluation in each iteration.}

\cref{fig:type-dist} shows the count of each semantic type in the dataset $\mathcal{D}$. The distribution is clearly unbalanced with a long tail. Single-column models tend to perform poorly on the less-common types that comprise the long-tail. By effectively incorporating context, \sato significantly improves prediction accuracy for those types.

To better understand relationships between the  semantic types of columns in the same table, we conduct a preliminary analysis on the co-occurrence patterns of types. \cref{fig:co-occur}, shown in log-scale for readability, reports the frequencies of selected pairs of semantic types occurring in the same table. Most frequently co-occurring pairs include (\semantic{city}, \semantic{state}), (\semantic{age}, \semantic{weight}), (\semantic{age}, \semantic{name}), (\semantic{code}, \semantic{description}).

\subsection{Feature extraction}
We use the public Sherlock feature extractors\footnote{\url{https://github.com/mitmedialab/sherlock-project}} to extract the four groups of base features,  \charfeature, \wordfeature, \parafeature and \statfeature, \new{generating a feature vector with 1587 dimensions for each column in a table. }
\new{Those features have been proven effective for semantic type detection and provide good coverage of the granularity spectrum, ranging from character-level distribution features to global statistics. In addition, the \wordfeature and \parafeature features take advantage of powerful pre-trained word and paragraph embeddings which enable a better understanding of natural language contents.}

To make a fair comparison, these {\it base} features were used by both baseline methods and proposed methods in the experiments.
To generate table topics as introduced in Section~\ref{sec:model-LDA}, we train an LDA model that captures the mapping from \tableVocab to the latent topic dimensions. Since LDA is an unsupervised model,  we only need the vocabulary (i.e., set of all cell values) of the tables without any headers or semantic annotation. We convert numerical values into strings and then concatenate all values in the table sequentially to form a ``document'' for each table. Using the gensim~\cite{rehurek_lrec} library, we train an LDA model with 400 topics on a separate dataset of 10K tables. With the pre-trained LDA, we can extract topic vectors for tables using values from the entire table as input. Every table has a single topic vector, shared across columns. 

\subsection{Model implementation}\label{subsec:implementation}
We implement the multi-input neural network introduced in \cite{Hulsebos:2019:KDD} using PyTorch \cite{paszke2017automatic} as the \base single-column model. Throughout the experiments discussed here, we train the \base neural network model for 100 epochs using the Adam optimizer with a learning rate of $1e-4$ and a weight decay rate of $1e-4$.

For topic-aware prediction in \sato, the table topic features go through a separate subnetwork with an architecture identical to the subnetworks of the \base feature groups. Before going into the primary network, the outputs of all four subnetworks are concatenated with \statfeature to form a single vector. We train \sato's CRF layer with a batch size of 10 tables, using the Adam optimizer with a learning rate of $1e-2$ for 15 epochs. We initialize the pairwise potential parameters of the CRF model with the column co-occurrence matrix calculated from a held-out set of the WebTables corpus. We set the CRF unary potentials for columns to be their normalized topic-aware prediction score.

\subsection{Evaluation metrics}
    We measure the prediction performance on each target semantic type by calculating $\text{F}_{1}=2\times\frac{\text{precision} \times \text{recall}} {\text{precision} + \text{recall}}$. Since the semantic type distribution is not uniform, we report two types of average performances using the \sFone and \mFone.  The \sFone score is the average of per-type $\text{F}_{1}$ values weighted by support (sample size in the test set for the respective type) and reflects the overall performance. The \mFone score is the unweighted average of the per-type $\text{F}_{1}$ scores, treating all types equally, and is therefore more sensitive to types with small sample sizes compared to \sFone. 

\begin{table*}[ht]
\centering
\begin{tabular}{@{}lllll@{}}
\toprule
         & \multicolumn{2}{c}{Multi-column tables $\mathcal{D}_{mult}$}          & \multicolumn{2}{c}{All tables $\mathcal{D}$}                  \\ 
         & \MFone                  & \SFone                 & \MFone                 & \SFone                 \\ \midrule
\base    & 0.642 $\mscript{\pm 0.015}$                                   & 0.879 $\mscript{\pm 0.002}$                            & 0.692 $\mscript{\pm 0.007}$                            & 0.867 $\mscript{\pm 0.003}$           \\
\sato    & \textbf{0.735} $\mscript{\pm 0.022}$ (14.4\%$\uparrow$)       & \textbf{0.925} $\mscript{\pm 0.003}$ (5.3\%$\uparrow$) & \textbf{0.756} $\mscript{\pm 0.011}$ (9.3\%$\uparrow$) & \textbf{0.902} $\mscript{\pm 0.002}$ (4.0\%$\uparrow$) \\
\satoLDA & 0.713 $\mscript{\pm 0.025}$ (11.0\%$\uparrow$)                & 0.909 $\mscript{\pm 0.002}$ (3.5\%$\uparrow$)          & 0.746 $\mscript{\pm 0.011}$ (7.8\%$\uparrow$)          & 0.891 $\mscript{\pm 0.003}$ (2.8\%$\uparrow$)          \\
\satoCRF & 0.681 $\mscript{\pm 0.016}$ (6.6\%$\uparrow$)                 & 0.907 $\mscript{\pm 0.002}$ (3.2\%$\uparrow$)          & 0.711 $\mscript{\pm 0.006}$ (2.9\%$\uparrow$)          & 0.884 $\mscript{\pm 0.002}$ (2.0\%$\uparrow$)          \\ \bottomrule
\end{tabular}
\caption{
Performance comparison of the methods across the datasets $\mathcal{D}_{mult}$ (multi-column only) and $\mathcal{D}$ (the full dataset)
\new{Numbers are the average values over a 5-fold cross validation. $\pm$ denotes $95\%$ CI. () shows the relative improvements in percentage over \base. We conducted statistical tests using paired $t$-test with Bonferroni correction for multiple comparisons. \sato, \satoLDA, \satoCRF perform significantly better than \base ($p<.005$ in all metrics.) \sato performs significantly better than \satoLDA ($p<.005$ in all metrics) and \satoCRF ($p<.005$ on $\mathcal{D}_{mult}$ and n.s. on $\mathcal{D}$.)}
}\label{tab:overall-res}
\end{table*}

\section{Results}\label{sec:results}

\cref{tab:overall-res} reports improvements of the \sato variants over the \base method on both the dataset $\mathcal{D}_{mult}$, which includes only tables with more than one column, and the complete dataset $\mathcal{D}$. \new{We implemented \base using features and neural network structure of the Sherlock \cite{Hulsebos:2019:KDD} model.} 
On multi-column tables, \sato improves the \mFone score by \new{0.093 ($14.4\%$)} and the \sFone score by \new{0.046 ($5.3\%$)} compared to the single-column \base. When evaluated on all tables we still see a 0.064 ($9.3\%$) improvement on \mFone score and 0.035 ($4.0\%$) improvement on \sFone, although these scores are diluted by the inclusion of tables without valid table context. The results confirm that \sato can effectively improve the accuracy of semantic type prediction by incorporating contextual information embedded in table semantics.
%

We also evaluate the variants of \sato with single components: \satoLDA only performed topic-aware prediction using \tableVocab and \satoCRF conducted structured prediction using \base output as unary potential without considering table topic features. As shown in \cref{tab:overall-res}, both \satoLDA and \satoCRF provide improvements over the \base model but are outperformed by the combined effort in \sato. The results indicate that the structured prediction model and the topic-aware prediction model make use of different pieces of table context information for semantic type detection.

We note that there are always larger improvements on \mFone scores than \sFone scores, suggesting that a significant amount of \sato's improvements come from boosting accuracy for the less represented types. To better understand the influence of techniques used in \sato, we next perform a per-type evaluation for both \sato components on multi-column tables.

\subsection{Topic-aware prediction}

\begin{figure*}[ht]
\begin{subfigure}[l]{\linewidth}
\centering
\includegraphics[width=\textwidth]{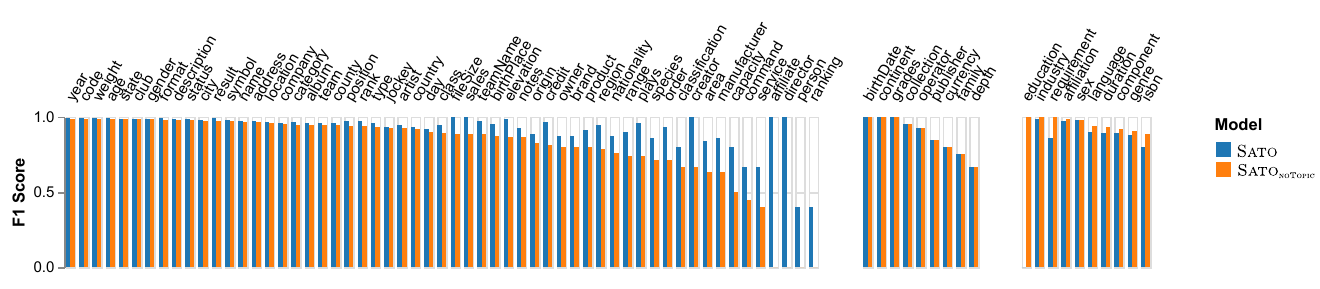}
\caption{\sato vs. \satoCRF \label{fig:SATO-SATO_CRF}}
\end{subfigure}%
\\
\begin{subfigure}[l]{\linewidth}
\centering
\includegraphics[width=\textwidth]{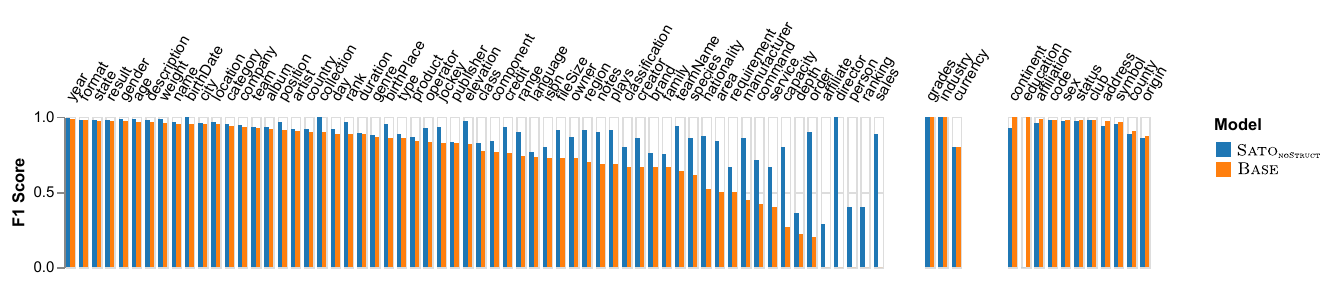}
\caption{\satoLDA vs. \base \label{fig:SATO_LDA-base}}
\end{subfigure}
 \caption{\Fone scores for each type obtained with (blue) and without (orange) topic-aware prediction. (a) compares \sato and \satoCRF (\sato without the topic-aware module), (b) compares \satoLDA (\base with topic) and \base, showing improvements on the majority of types. The effect is significant for many underrepresented types. \label{fig:per-type-topic}}
\end{figure*}

\cref{fig:per-type-topic} shows the per-type comparison of \Fone scores between models with and without the topic-aware prediction component. More specifically, \cref{fig:SATO-SATO_CRF} compares the full \sato against \sato without \tableVocab (i.e., \satoCRF,) and \cref{fig:SATO_LDA-base} compares \satoLDA (only topic-aware model) against \base. Including information in \tableVocab improved 59 out of 78 semantic types for \satoCRF with 9 types getting equal and 10 types getting worse performances. Similarly, \satoLDA improves the performance for 64 types and decreases it for 11 types. The prediction performance stays unchanged for 3 types.
 
We also see significant improvements in the previously ``hard'' semantic types with small support size. The types with the highest accuracy increases, \semantic{affiliate}, \semantic{direc\-tor}, \semantic{person}, \semantic{ranking}, and \semantic{sales},  all come from the fifteen least represented types as shown in \cref{fig:type-dist}. This shows incorporating \tableVocab effectively alleviates the problem of lacking training data for the rare types. 

\begin{figure*}[ht]
\begin{subfigure}[l]{\linewidth}
\centering
\includegraphics[width=\textwidth]{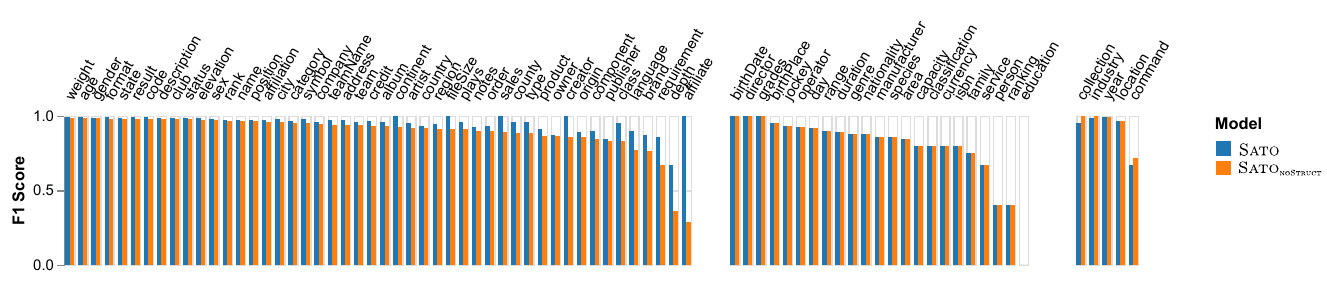}
\caption{\sato vs. \satoLDA \label{fig:SATO-SATO_LDA}}
\end{subfigure}%
\\
\begin{subfigure}[l]{\linewidth}
\centering
\includegraphics[width=\textwidth]{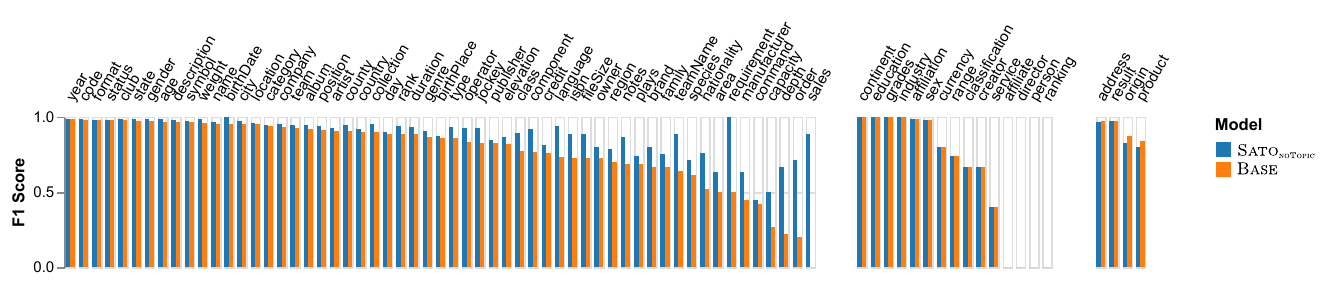}
\caption{\satoCRF vs. \base \label{fig:SATO_CRF-base}}
\end{subfigure}
\caption{\Fone scores for each type obtained with (blue) and without (orange) structured prediction (a) compares \sato and \satoLDA (\sato without the structured prediction module), (b) compares \satoCRF (\base with structured prediction) and \base, showing improvements on the majority of types. Although the improvements on long-tail types are less significant compared to the topic-aware model in \cref{fig:per-type-topic}, fewer types get worse predictions (shown in the right panels). Structured prediction can correct mispredictions  by directly modeling column relationships. \label{fig:per-type-CRF}}
\end{figure*}

\subsection{Structured prediction}
To evaluate the contribution of structured prediction, we compare \sato with its variant without structured prediction, \satoLDA (\cref{fig:SATO-SATO_LDA}).  Similarly, we compare the performance 
of \satoCRF (structured prediction directly on \base output) with that of \base (\cref{fig:SATO_CRF-base}). \base is improved on 50 types and \satoLDA is improved on 59 types with structured prediction. For a subset of rare types (e.g., {\tt depth}, {\tt sales}, {\tt affiliate},) the prediction accuracy is dramatically improved. While for others (e.g., {\tt person}, {\tt director}, {\tt ranking},) there is no noticeable improvement as with topic-aware prediction. This shows structured prediction is less effective in boosting the accuracy of rare types compared to topic-aware prediction. However, at the same time, both the number of types that get worse accuracy (4 and 5 respectively) and the drop in \Fone scores for those types are smaller with structured prediction as compared to topic-aware prediction. Enforcing table-level context can be too aggressive sometimes, leading to worse performance for certain types. Through modeling relationships between inferred types of surrounding columns, the structured prediction module in \sato ``salvages'' some of these overly aggressive predictions. We conduct qualitative analysis in \cref{sec:qualitative} to further look into this effect.


In conclusion, multi-column predictions from the structured prediction model, with or without topic modeling, outperforms the column-wise models. 

\begin{figure}[h]
    \centering
    \includegraphics[width=0.95\columnwidth]{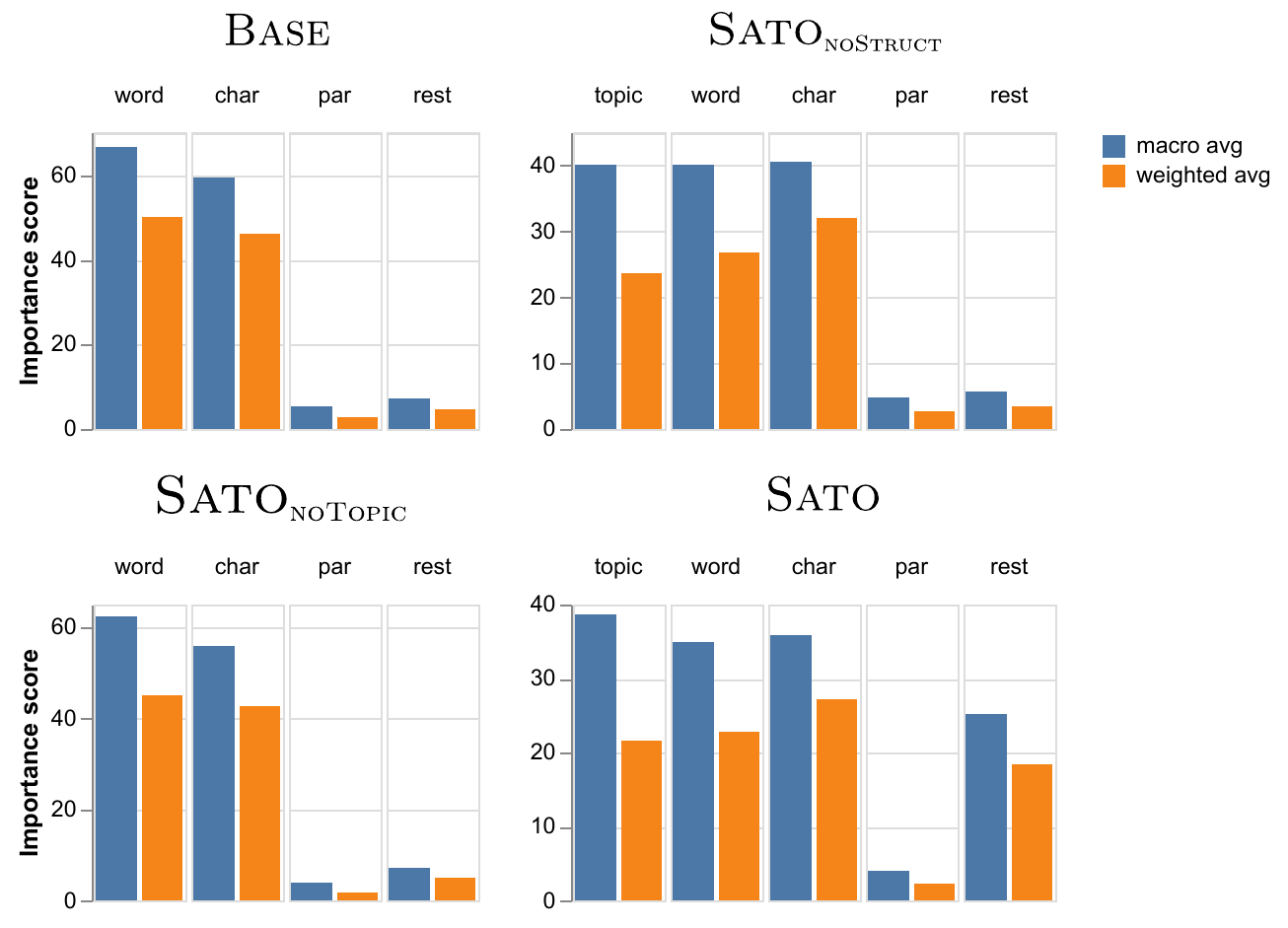}
    \caption{Importance scores for the feature categories used in our models obtained by measuring the drop in both aggregated \Fone values from permutation experiments. 
    {\it Topic} features are the most important feature category with respect to the \mFone score in the full \sato model, providing additional evidence for the contribution of topic modeling in predicting underrepresented semantic types. 
    \label{fig:FI}
    }
\end{figure}

\subsection{\new{Efficiency}}
\label{sec:efficiency}
\new{
We show that the \sato model successfully improves prediction accuracy by introducing the topic-aware features and the CRF layer. However, the additional components may cause additional time cost. To evaluate the efficiency of \sato, we repeated the training and prediction procedures for 5 times and measured the training and prediction time of \base and \sato on the multi-column dataset $\mathcal{D}_{mult}$. The training data contains 26K tables and the test data contains 6.4K tables. For further investigation on the cost of the topic-aware features and the CRF layer, we separately measured the time for training the main model, and the time for training the CRF layer. We use the same hyper-parameters used in the experiment (described in \ref{subsec:implementation}) for both of the models for a fair comparison. 
The experiment was conducted on a single machine with 2.1GHz CPUs (64 cores) and 512GB RAM.
\cref{table: efficiency} summarizes the average training and prediction time for those models. 
%
}

\new{
From the results, we confirm that adding the topic-aware features and the CRF layer increases approximately 81 s and 367 s for training time, respectively. We would like to emphasize that we do not need to retrain a model unless we obtain a significant amount of additional training data. Thus, we consider that the difference is not critical. On average, \sato takes +1.4 s than \base to generate predictions for all 6.4K tables in the test set of $\mathcal{D}_{mult}$, which is 0.2 ms per table.
%
We believe the overhead will be mostly unnoticeable in practice, and the average prediction time per table (0.8 ms) is can support the interactive use of \sato.
}


\begin{table}[t]
\centering
\begin{tabular}{@{}lccc@{}}
\toprule
         & \multicolumn{2}{c}{Training time [s]} & Prediction time [s] \\
          & \begin{tabular}[c]{@{}l@{}}Features\end{tabular} & \begin{tabular}[c]{@{}l@{}}Structured\end{tabular} &  \\         
         \midrule

\base    & $596.9 \pm 9.2$                                             & N/A                                                            & $3.8 \pm 0.04$        \\
\sato    & $678.5 \pm 15.1$                                            & $366.9 \pm 66.8$                                           & $5.2 \pm 0.06$       \\
\bottomrule
\end{tabular}
\caption{\new{Average training and prediction time over 5 trials $\mathcal{D}_{mult}$. $\pm$ denotes $95\%$ CI. Training time for the column-wise features (Features) and the CRF layer (Structured) is reported separately.}}
\label{table: efficiency}
\vspace{-6mm}
\end{table}

\subsection{Feature importance}
To better understand the influence of the different feature groups, we perform permutation importance \cite{altmann2010permutation} analysis on \base and \sato variants. For each fitted model and a specific feature group, we take the input tables and perform shuffling by only swapping features in the specified feature group with randomly selected tables. Such feature mismatch will cause less accurate predictions and a worse overall performance. Shuffling crucial features will break the strong relationships between input and output, leading to a significant drop in accuracy. We took the average of the normalized drop in \Fone scores over five random trials as the feature importance measurement. 

\cref{fig:FI} shows that for both the \base model and \satoCRF, the \wordfeature and \charfeature feature groups are the most important feature groups. This matches the conclusions in \cite{Hulsebos:2019:KDD}. When considering the global context, the additional {\it Topic} feature group has comparable or greater importance than \wordfeature and 
\charfeature. The effect is more obvious with respect to the \mFone metric, confirming the help of \tableVocab information, especially on less-represented types.

\subsection{Topic interpretation}\label{sec:topic_analysis}
We conduct qualitative analysis on the LDA model to 
investigate how the model captures semantics from each table and provides contextual information to \NAME. To obtain the topic distribution of each semantic type, we calculate the average topic distribution based on the topic distributions $\theta_i$ of the $i$-th table that contains the semantic type. For each topic,  we chose top-$k$ semantic types as representative semantic types by the probability of the topic. 

We find that some topics had ``flat'' distributions where most semantic types have almost the same probabilities. Since these topics are not very useful for classifying semantic types, we compute a saliency score for each topic and sort the topics by their saliency.  Our saliency score averages the probabilities of the top-$k$ semantic types for each topic.

\cref{tab:lda_topic_analysis} shows the top-5 salient topics and the representative semantic types. Following the standard approach in topic model analysis~\cite{Blei:2012:TopicModels, Blei:2003:LDA}, we manually devise an interpretation for each topic. For example, topic dimension \#192 and \#99 are activated by personal information in table values, whereas \#264 is closely related to business tables. These examples demonstrate that semantic space learned using LDA could capture intent information from tables.

\begin{table}[]
\begin{tabular}{@{}lll@{}}
\toprule
Topic & Top-5 semantic types                                                                      & Interpretation       \\ \midrule
192   & \begin{tabular}[t]{@{}l@{}}origin, nationality, country,\\ continent, sex\end{tabular}    & person               \\
99    & \begin{tabular}[t]{@{}l@{}}affiliate, class, person, \\ notes, language\end{tabular}      & person               \\
394   & \begin{tabular}[t]{@{}l@{}}religion, family, address, \\ teamName, publisher\end{tabular} & person, book         \\
264   & \begin{tabular}[t]{@{}l@{}}code, description, creator,\\ company, symbol\end{tabular}     & business             \\ \bottomrule
\end{tabular}
\caption{Examples of the topics learned by the LDA model, 
semantic types associated with each topic that are obtained 
by using a saliency metric, and our interpretation for each topic. \label{tab:lda_topic_analysis}}
\end{table}

\subsection{Column embeddings (Col2Vec)}
To verify how the table intent features help the \NAME{} model capture the table semantics, we analyze and compare the embedding vectors from the final layer of the \NAME{} model and the baseline Sherlock model as {\it column embeddings.} We can consider these embeddings as {\it column embeddings} since the final layer combines input signals to compose semantic representations. For comparison, we used the final layer of the single-column prediction model of \NAME{}, before the CRF layer. Therefore, we assume that the Table Intent features account for the difference in the embeddings.

\begin{figure}
    \centering
	\begin{subfigure}[b]{0.425\linewidth}
	\centering
	\includegraphics[width=0.98\textwidth]{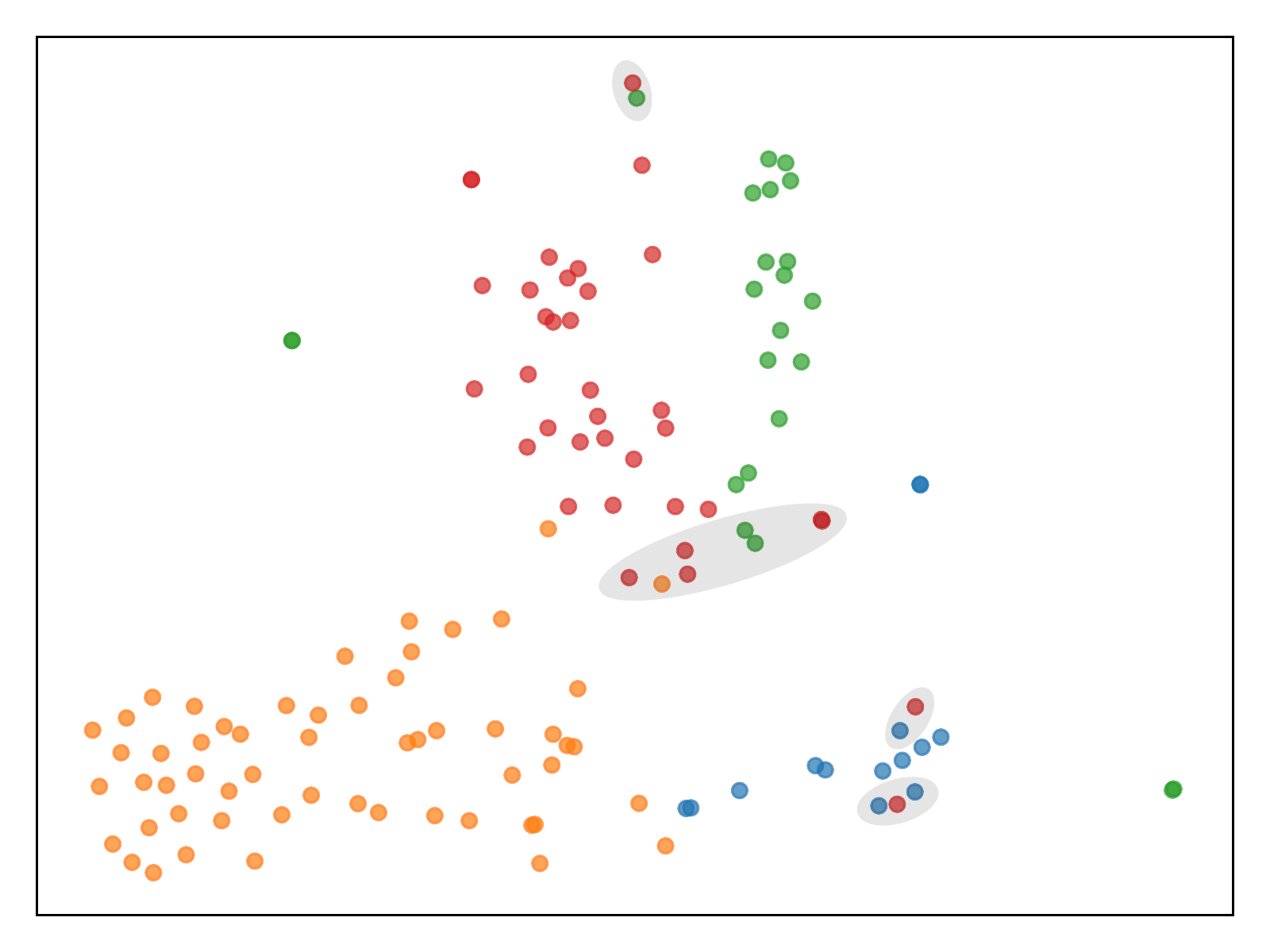}
	\caption{\label{fig:emb_sherlock}}
	\end{subfigure}%
	\begin{subfigure}[b]{0.425\linewidth}
	\centering
	\includegraphics[width=0.98\textwidth]{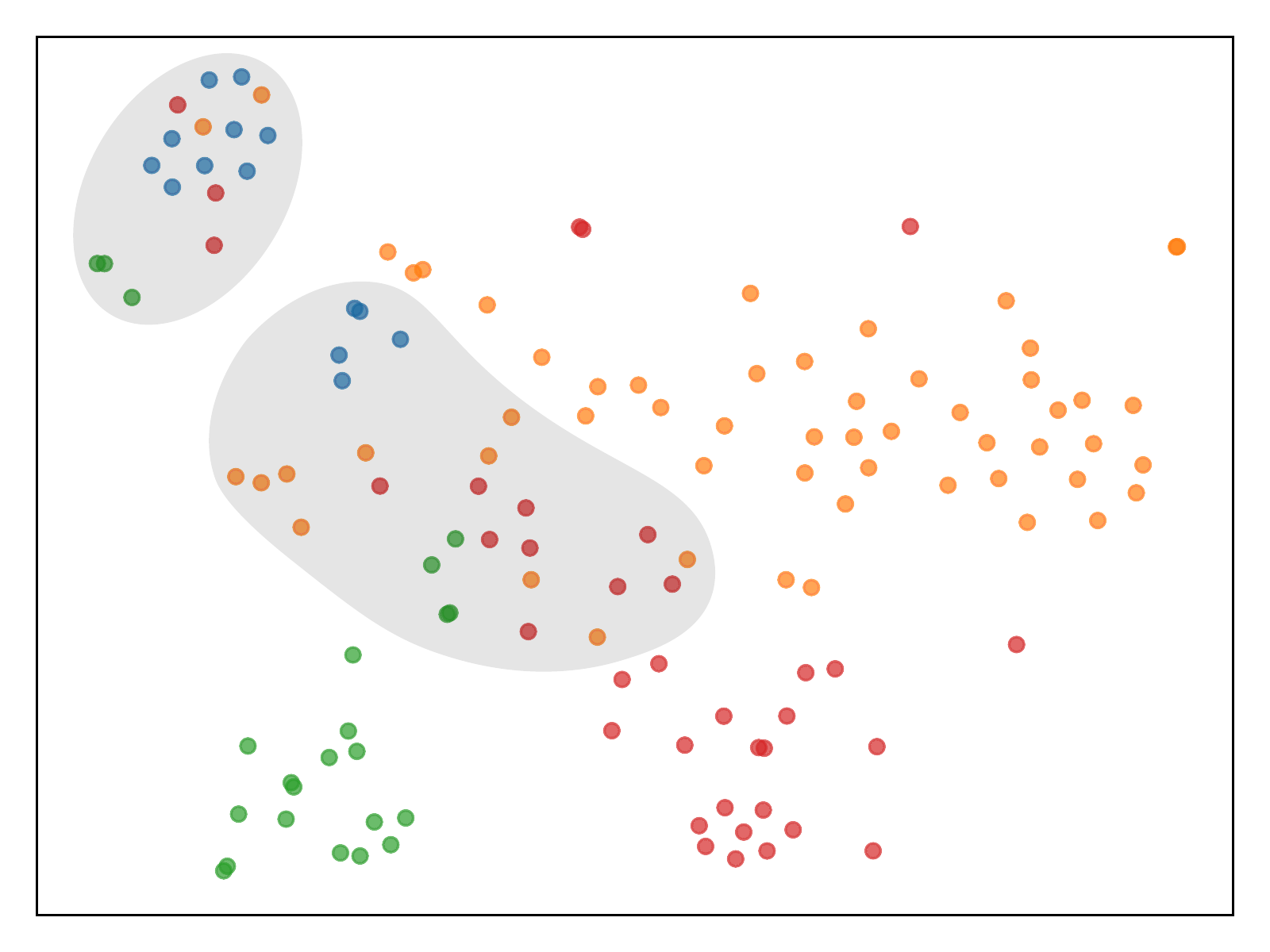}
	\caption{\label{fig:emb_sherlock_lda}}
	\end{subfigure}
    \begin{subfigure}[b]{0.15\linewidth}
	\centering
	\includegraphics[width=0.98\textwidth]{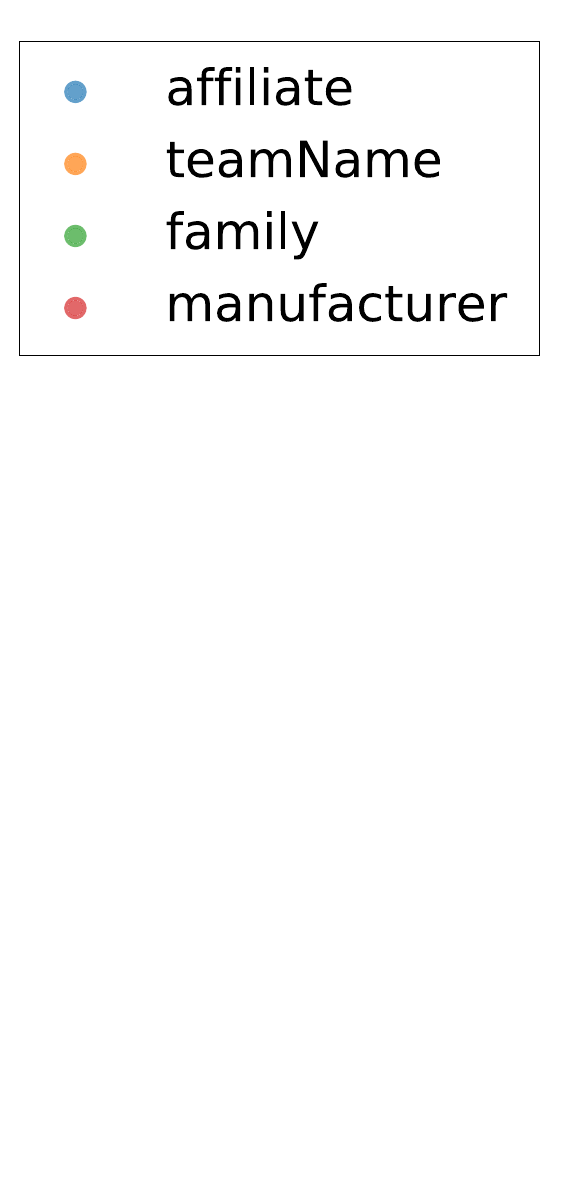}
    \end{subfigure}
\caption{
Two-dimensional visualizations of column embeddings by (a) \satoLDA, and (b) \SHERLOCK. Colors denote semantic types. Gray-colored regions are manually added to emphasize the areas of ``ambiguity''in the  column embeddings. \satoLDA appears to separate similar semantic types better. \label{fig:embedding_analysis}
 \vspace{-3mm}
}
\end{figure}

Following prior examples (e.g.,~\cite{Zeiler:2014:Visualizing}), we analyze column embeddings of the test columns used in the experiments. We use t-SNE~\cite{vanDerMaaten:2008:tSNE} to reduce the dimensionality of the embedding vectors to two and then visualize them using a two-dimensional scatterplot. To embed vectors of the two methods in a common space, we fit a single t-SNE model for all data points, and then visualize major semantic types that are related to organizations ({\tt affiliate}, {\tt teamName}, {\tt family}, and {\tt manufacturer}) to investigate how the \NAME{} model with the Table Intent features can distinguish columns of those ambiguous semantic types. 

\cref{fig:embedding_analysis} shows the visualization of embedding vectors of \NAME{} and Sherlock. With Sherlock, the column embeddings of each semantic type partially form a cluster, but some clusters are overlapped compared to the column embeddings by \NAME{}. In \cref{fig:embedding_analysis} (a), we observe a clearer separation between the organization-related semantic types with little perturbation. The results qualitatively confirm that topic-aware prediction helps \NAME{} distinguish semantically similar semantic types by capturing the table context of an input table.  Note that these column embeddings are from the test set, and any label information from these columns was not used to obtain the column embeddings. Thus, we can also confirm that \NAME{} appropriately generalizes and learns column embeddings for these semantic types.

\vspace{5pt}
\subsection{Qualitative analysis}\label{sec:qualitative}

\begin{table*}[ht]
\centering
\begin{subtable}{\linewidth}
\centering
\begin{tabular}{@{}llll@{}}
    \toprule
    Table ID & True Columns                 & \base (w/o structured prediction)            & \satoCRF (w/ structured prediction)                \\ \midrule
    6299     & code, name, city             & \mispred{symbol}, \mispred{team}, city           & \crtpred{code}, \crtpred{name}, city             \\
    898      & company, location            & \mispred{name}, \mispred{city}                   & \crtpred{company}, \crtpred{location}            \\
    4575     & symbol, company, isbn, sales & symbol, \mispred{name}, isbn, \mispred{duration} & symbol, \crtpred{company}, isbn, \crtpred{sales} \\
    5712     & type, description            & \mispred{weight}, \mispred{name}                 & \crtpred{type}, \crtpred{description}            \\
    3865     & year, teamName, age          & year, \mispred{city}, \mispred{weight}           & year, \crtpred{teamName}, \crtpred{age}          \\ \bottomrule
    \end{tabular}
    \caption{Corrected tables from \base predictions\label{tab:qualitative-base}}
\end{subtable}

\begin{subtable}{\linewidth}
\centering
    \begin{tabular}{@{}llll@{}}
    \toprule
    Table ID & True Columns             & \satoLDA (w/o structured prediction)                    & \sato (w/ structured prediction)                    \\ \midrule
    410      & brand, weight            & \mispred{artist}, \mispred{code}                & \crtpred{brand}, \crtpred{weight}           \\
    5655     & code, name, city         & \mispred{club}, name, \mispred{name}            & \crtpred{code}, name, \crtpred{city}         \\
    4369     & day, location, notes     & \mispred{name}, location, \mispred{location}    & \mispred{name}, location, \crtpred{notes}    \\
    30       & language, name, origin   & language, name, \mispred{description}           & language, name, \crtpred{origin}   \\
    4531     & rank, name, city         & rank, \mispred{location}, \mispred{location}    & rank, \mispred{location}, \crtpred{city}    \\
    \bottomrule
    \end{tabular}
\caption{Corrected tables from \satoLDA predictions\label{tab:qualitative-LDAbase}}
\end{subtable}
\caption{Examples of the  mispredictions that are corrected by performing a structured prediction using the linear-chain CRF.\label{tab:qualitative}}
\end{table*}

To better understand how structured prediction further helped \sato with the existence of topic-aware predictions, we conducted qualitative analysis by identifying examples where table-wise prediction ``salvages'' bad predictions in the column-wise (i.e., \base and \satoLDA) predictions.

\cref{tab:qualitative-base} shows a selected set of example tables from the test sets where the  incorrect predictions from the \base model are corrected by applying structured prediction using our trained CRF layer. For example, with table \#4575, the columns \semantic{company} and \semantic{sales} were incorrectly predicted as \semantic{name} and \semantic{duration} by the single-column \base model. By modeling inter-column dependencies, \satoCRF correctly predicts  the types \semantic{company} and \semantic{sales},  which tend to co-occur more with surrounding columns \semantic{symbol} and \semantic{isbn} for tables about books and magazines. \cref{tab:qualitative-LDAbase} shows examples where \satoLDA made incorrect predictions using \tableVocab and was subsequently corrected by the use of structured prediction (i.e., \sato). Table \#4369 and table \#4531 are examples where location-related vocabulary in tables made a large impact. It produced overly aggressive predictions with multiple \semantic{location} columns, whereas \sato with the additional structured inference step successfully corrected one of the columns.

Furthermore, considering surrounding types, structured prediction effectively improves performance for numerical columns like \semantic{duration/sales} from table \#4575, \semantic{age/weight} from table \#3865, \semantic{code/weight} from table \#410. 

\section{Discussion}\label{sec:discussion}

\subhead{Using learned representations}\label{sec:learned-rep}
\sato's single column prediction module based on Sherlock incorporates four categories of features that characterize different aspects of column values, amassing more than 1.5K feature values per column. However, the availability of large-scale table corpora presents a unique  opportunity to develop pre-trained representation models and eschew manual feature extraction. 
To test the viability of using representation models, we fine-tuned the BERT model~\cite{devlin2019bert}, a state-of-the-art model for language representation, for our semantic type detection task. Models based on fine-tuning BERT have recently improved  prior 
art on several NLP benchmarks without manual featurization~\cite{devlin2019bert,liu2019fine,liu2019roberta}.  We trained the BERT model using the default BERT parameters, achieving a support-weighted F1 score of 0.866, which is slightly better than 0.852 achieved by the Sherlock model. This  result is promising because a ``featurization-free'' method with default parameters is able to achieve a prediction accuracy comparable to that of Sherlock. However, our multi-column prediction still outperforms the BERT model by a large margin, indicating the importance of incorporating table context into column type prediction. A promising avenue of future research is to combine our multi-column model with BERT-like pre-trained learned representation models.
\\
\\
\subhead{Exploiting type hierarchy through ontology}
In this paper, we consider semantic types without hierarchy. \shepherding{However, it is possible to form natural parent-child relationships between many types. For instance, {\tt country} and {\tt city} are types (subclasses) of {\tt location} and {\tt club} and {\tt company} are types of {\tt organization}. Factoring  hierarchical type relations into prediction (e.g.,~\cite{limaye2010annotating,Takeoka:2019:Meimei}) requires an ontology codifying the type hierarchy and, crucially, additional annotation over training dataset, which can be infeasible to manually carry out for large training datasets such as the one used here.  Nevertheless, modeling and predicting hierarchical semantic types can provide richer information for downstream tasks. It can also further improve the prediction accuracy by leveraging the additional structure afforded by hierarchical relations,  especially for the types that have small numbers of training samples.} 

\smallskip
\subhead{\new{High-order CRFs}} \label{sec:high_order_CRF}
\new{
Several studies~\cite{Cuong:2014:JMLR:HighorderCRF,Lavergne:2017:VariableOrderCRF,Qian:2009:HigherOrderCRF} developed high-order CRF models that implement potential functions that take $n$ ($n>2$) predictions into account. 
However, the computational complexity of exact inference steps for training and prediction becomes exponentially expensive: $\mathcal{O}(L^K)$, where $L$ is the input sequence length  (i.e., \# of columns) and $K$ is the number of states (i.e., \# of semantic types.) The computational cost is significantly expensive compared to the original linear-chain CRF’s $\mathcal{O}(K L^2)$.
As \sato with the linear-chain CRF model significantly improved the performance for the semantic type detection task, we decided not to use the degree of the order for efficiency.}

\new{
Additionally, we believe that high-order dependency between predictions is not always necessary if we incorporate contextual features into the model. 
\cite{Qian:2009:HigherOrderCRF} shows that contextual  features that take into account surrounding information are more useful than a high-order CRF architecture for named entity recognition tasks. Since table topic features provide table-wise contextual information, we consider the original CRF model with pairwise potential functions as the right choice for improving the model accuracy efficiently.}


\vspace{-3mm}
\section{Related Work\label{sec:related}}

\subhead{Regular expression and dictionary lookup}
Semantic type detection enhances the functionality of commercial data preparation and analysis systems such as Microsoft Power BI \cite{powerbi}, Trifacta \cite{trifacta}, and Google Data Studio \cite{googledatastudio}.
These commercial tools typically rely on manually defined rule-based approaches such as regular expression patterns dictionary lookups to detect semantic types. For instance, Trifacta detects around 10 types and Power BI only supports time-related semantic types. Open source libraries such as messytables~\cite{messytables}, and csvkit~\cite{csvkit} similarly use heuristics to detect a limited set of types.
\\
\\
\subhead{Ontology-based}
Prior work, with roots in the semantic web and schema matching literature, provide alternative approaches to semantic type detection. One body of work leverages existing data on the web, such as WebTables \cite{webtables}, and ontologies (or, knowledge bases) such as DBPedia \cite{dbpedia}, Wikitology \cite{syed2010exploiting}, and Freebase~\cite{freebase}. Venetis et al. \cite{venetis2011recovering} construct a database of value-type mappings, then assign types using a maximum likelihood estimator based on column values. Syed et al. \cite{syed2010exploiting} use column headers and values to build a Wikitology query mapping columns to types. 
\\
\\
\subhead{Statistical similarity}
Several earlier approaches rely on statistical similarity or other measures of data similarity to match columns with types. Ramnandan et al.~\cite{ramnandan2015assigning} first separate numerical and textual column types, then compare column values to those with labels from a dataset using the Kolmogorov-Smirnov (K-S) test and Term Frequency-Inverse Document Frequency (TF-IDF,) respectively. Pham et al.~\cite{pham2016semantic} use additional features and tests, including the Mann-Whitney test for numerical data and Jaccard similarity for textual data, to train logistic regression and random forest models. 
\\
\\
\subhead{Synthesized}
Puranik~\cite{puranik} proposes combining the predictions of ``experts,'' including regular expressions, dictionaries, and machine learning models. More recently, Yan and He \cite{yan2018synthesizing} introduced a system that, given a search keyword and a set of positive examples, synthesizes type detection logic from open source GitHub repositories. This system provides a novel approach to leveraging domain-specific heuristics for parsing, validating, and transforming semantic types. 
\\
\\
\subhead{Learned}
\shepherding{Another line of prior work employs machine learning, including probabilistic graphical models. Goel et al.~\cite{goel2012exploiting} split each cell (field) value in a table into tokens and attempted to predict the field and token labels using CRF models with different graph structures capturing dependencies among tokens and fields.  For instance, a cell value  `Mountain View, CA'  is split into a sequence of tokens `Mountain', `View',`,', `CA'. Then a multi-layer CRF model is used to assign labels {\tt cityName}, {\tt cityName}, {\tt symbol}, and {\tt state} for those tokens along with the cell label {\tt place}. This approach requires curating cell- and token-level annotations for training, which is impractical for large-scale table corpora. Furthermore, it has limited robustness over missing, dirty, and heterogeneous data, as well as semantic data types with highly variable formatting.  \sato avoids the need for fine-grained token-level annotations and only uses automatically annotated column level semantic types, relying on the 
power of a deep neural network and word embeddings to capture cell-level information.}

\shepherding{Limaye et al.~\cite{limaye2010annotating}  use a Markov random field (MRF) model to annotate values with entities, columns with types, and column pairs with relationships. This approach assumes the existence of a catalog specifying entities, types, and relations between them and relies on good matches between entity lemma and cell text to make accurate predictions of both cell and column types. However, in practice, an accurate catalog can be expensive or impossible to obtain for large corpora or new domains and many tables have missing or noisy (incomprehensible, malformed, etc.) headers. Takeoka et al.~\cite{Takeoka:2019:Meimei}  extend Limaye et al.~\cite{limaye2010annotating}'s work with multi-label classifiers to support additional types, including numerical data types, and improve its predictive performance. However, this approach also relies on training data (183 tables) collected through human annotation and its application to massive table corpora can get extremely expensive.}

\shepherding{Similar to earlier approaches~\cite{goel2012exploiting,limaye2010annotating,Takeoka:2019:Meimei} discussed above, \sato also uses a probabilistic graphical model for structured output prediction. However, in contrast to this earlier work, \sato employs a CRF model to combine the topic-aware predictions of a large-scale deep learning model, leveraging a large number of real-world tables for training. These tables are automatically annotated  without resorting to human labeling, which makes \sato easier to extend and scale than prior work using probabilistic graphical models.}

Although prior research used shallow neural networks for related tasks (e.g.,~\cite{li1994semantic}),   Sherlock~\cite{Hulsebos:2019:KDD} is the first deep learning model directly applied to semantic type detection for table columns.  Trained on a large number of columns, Sherlock uses a multi-input 
neural network to make type prediction based on features of column 
values. \sato builds on Sherlock and addresses its two related 
drawbacks; the low prediction accuracy for underrepresented types and the lack of consideration for table context in prediction. 

\vspace{-3mm}
\section{Conclusion\label{sec:conclusion}}
 Automated semantic typing is becoming more important than ever due to a rapid increase in the demand for better 
 data preparation tools. The semantics of a table column (or any other data source for that matter) are embodied by its context as well as its raw data values. Here, we introduce \sato to automatically detect the semantic types of table columns, leveraging the signals from the table context of columns as well as the data values of columns. \sato combines the power of large-scale deep learning together with structured prediction and topic modeling to achieve a prediction performance that significantly exceeds the state-of-the-art. Through ablation and permutation experiments, we evaluate \sato extensively and show how individual modeling choices as well as feature types contribute to the performance. To facilitate future applications and extended research, we are publicly releasing our trained model and source code for training along with an interactive web application demonstrating \sato's use at~\url{https://github.com/megagonlabs/sato}. 
\balance

\vspace{-3mm}
\section{Acknowledgments\label{sec:ack}}
We thank Jonathan Engel for suggesting the name Sato and his proofreading help. We also thank Kevin Hu for his help in making the Sherlock source code accessible. 
\newpage
\bibliographystyle{abbrv}
\bibliography{main}  


\end{document}